\documentclass[reprint, amsmath, amssymb, aps, pra, showpacs]{revtex4-1}

\usepackage{graphicx}
\usepackage{dcolumn}
\usepackage{bm}

\begin{document}

\title{Efficient solver for a special class of convection-diffusion problems}

\author{Narain Karedla}
\email{narain.karedla@phys.uni-goettingen.de}
\affiliation{III. Institute of Physics, Georg August University, 37077 G\"ottingen, Germany.} 

\author{Jan Christoph Thiele}
\email{jan.christoph.thiele@phys.uni-goettingen.de}
\affiliation{III. Institute of Physics, Georg August University, 37077 G\"ottingen, Germany.} 

\author{Ingo Gregor}
\email{ingo.gregor@phys.uni-goettingen.de}
\affiliation{III. Institute of Physics, Georg August University, 37077 G\"ottingen, Germany.} 

\author{J\"org Enderlein}
\email{jenderl@gwdg.de}
\affiliation{III. Institute of Physics, Georg August University, 37077 G\"ottingen, Germany.} 

\date{\today}


\begin{abstract}
We describe an exact and highly efficient numerical algorithm for solving a special but important class of convection-diffusion equations. These equations occur in many problems in physics, chemistry, or biology, and they are usually hard to treat due to the presence of a convection term, represented by a first order derivative in the spatial co-ordinate. Our algorithm reduces the convection-diffusion equation to a pure diffusion problem within a complex-valued potential which can be solved efficiently and accurately using conventional parabolic partial differential equation solvers. 
\end{abstract}

\maketitle

\section{Introduction}

Convection-diffusion processes play an important role in physics, chemistry and biology. They are described by the convection-diffusion partial differential equation which models the transport of particles, heat, energy or other quantities in space over time, involving both directed transport (convection) as well as diffusive, stochastically driven transport. In statistical physics and probability theory, convection-diffusion equations are also known as Fokker-Planck equations and describe the spatio-temporal transport of a probability density within an external force field (convection). Thus, convection-diffusion equations are omnipresent, and their numerical solution is of great importance in diverse fields of science. 

However, accurate numerical integration of convection-diffusion equations is a challenging task, due to the presence of the convection term. Many sophisticated procedures and algorithms have been developed in the past, such as finite-difference methods \cite{DeCampos2014}, lattice Boltzmann methods \cite{Chen1998a}, or relating the problem to a Smoluchowski-Langevin stochastic differential equation \cite{Szymczak2003}, for which powerful numerical solvers are available. Another noteworthy method are direct front-tracking methods such as the level-set algorithm \cite{glimm2001conservative} that is designed to track even steep concentration gradients with high accuracy, but also thee methods can be computationally challenging. In a recent paper \cite{Karedla2014}, we presented a path-integral based approximate solution for some special convection-diffusion problems, but this approximation was still restricted to special geometries and parameters, and far from being generally applicable or being highly accurate. Here, we describe a much more general, efficient and accurate approach for numerically solving convection-diffusion problems \emph{where the flow velocity profile does not change along the flow direction}. Particular examples of this class of convection-diffusion problems are the famous Taylor-Aris flow in capillaries, or a shear-flow over a planar surface. In this manuscript, we present our general approach for modeling this class of diffusion-convection problems, and apply it to Taylor-Aries capillary flow, and to shear-flow  over an either absorbing or impermeable surface.

\section{Theoretical Background}

Let us consider the convection-diffusion equation 

\begin{align}
\frac{\partial c\left(\mathbf{r},t\right)}{\partial t} = D \Delta c\left(\mathbf{r},t\right) - v\left(\mathbf{r}_\perp\right) \partial_x c\left(\mathbf{r},t\right)
\label{eq:convdiff}
\end{align} 

\noindent where $c(\mathbf{r},t)$ is the local concentration of the transported and diffusing quantity at position $\mathbf{r}$ and time $t$, $D$ is the diffusion coefficient, and $v(\mathbf{r}_\perp)$ is a flow along the $x$-direction which depends only on the coordinates $\mathbf{r}_\perp$ perpendicular to $x$. This is the case for all laminar flows along straight channels with arbitrary cross-section and with arbitrary boundary conditions of the form $a \nabla_\perp c\left(\mathbf{r},t\right) + b c\left(\mathbf{r},t\right) = d$ with fixed constants $a, b$ and $d$, and where $\nabla_\perp$ denotes the gradient along $\mathbf{r}_\perp$. Two important examples of this kind which will be discussed in this paper are (i) convection-diffusion in a shear-flow over a surface, and (ii) convection-diffusion in a capillary flow. As already mentioned, the numerically difficult term is the convection term which involves spatial derivatives of the first order. In a recent paper \cite{Karedla2014}, we developed an approximate solution to this kind of equation by using  path-integral approach. From this effort, we have found that one can efficiently eliminate the convective term in the equation by representing the solution $c(\mathbf{r},t)$ in integral form as

\begin{equation}
c(\mathbf{r},t) = \int \frac{d k}{2 \pi} \tilde{w}(k, \mathbf{r}_\perp, t) \exp(i k x - D k^2 t) 
\label{eq:cint}
\end{equation}

\noindent where we have introduced the auxiliary function $\tilde{w}(k, \mathbf{r}_\perp, t)$. After inserting this expression into eq.~\eqref{eq:convdiff}, we find that this function has to obey the differential equation 

\begin{equation}
\frac{\partial \tilde{w}}{\partial t} = D \Delta_\perp \tilde{w} - i k v(\mathbf{r}_\perp) \tilde{w}
\label{eq:wtilde}
\end{equation}

\noindent where the operator $\Delta_\perp$ is the Laplacian along the transversal co-ordinates $\mathbf{r}_\perp$. The initial conditions for $\tilde{w}$ are 

\begin{equation}
\begin{split}
\tilde{w}(k, \mathbf{r}_\perp, t=0) = \tilde{c}_0(k, \mathbf{r}_\perp)
\label{eq:wtildeinitial}
\end{split}
\end{equation}

\noindent where $\tilde{c}_0(k, \mathbf{r}_\perp)$ is the Fourier transform along the $x$-co-ordinate of the initial concentration distribution $c_0(\mathbf{r})$:

\begin{equation}
\tilde{c}_0(k, \mathbf{r}_\perp) = \int_{-\infty}^\infty dx c_0(x, \mathbf{r}_\perp) \exp(-i k x)
\end{equation}

The differential equation (\ref{eq:wtilde}) describes a diffusion process within a complex-valued potential $i k v(\mathbf{r}_\perp)$. Such equations can be numerically solved in a very efficient way. First, one discretizes the lateral coordinates into an $N \times M$ grid, $\mathbf{r}_\perp \rightarrow \left\{x_j,y_k\right\}$ with $1 \leq j \leq N$ and $1 \leq k \leq M$, and one sorts the corresponding values $\tilde{w}_{j,k}=\tilde{w}(k, x_j, y_k, t)$ into a linear vector $\mathbf{\tilde{w}}$. Then, the discrete solution of eq.~\eqref{eq:wtilde} is given by the matrix equation 

\begin{equation}
\mathbf{\tilde{w}}(k,\mathbf{r}_\perp,t) = \exp\left[ t\left(D \hat{\boldsymbol{\Delta}}_\perp - i k \mathbf{\hat{V}} \right) \right] \cdot \mathbf{\tilde{w}}_0(k,\mathbf{r}_\perp)
\label{eq:wtildesol}
\end{equation}

\noindent where $\hat{\boldsymbol{\Delta}}_\perp$ is the $MN \times MN$ matrix of the discretized transversal Laplace operator corresponding to the ordering of the elements in $\mathbf{\tilde{w}}$ and encoding also the correct boundary conditions, and $\hat{\mathbf{V}}$ is an $MN \times MN$ diagonal matrix with elements $v(x_j,y_k)$. The exponential function on the r.h.s. in eq.~\eqref{eq:wtildesol} is understood as matrix exponentiation. Numerically, eq.~\eqref{eq:wtildesol} is efficiently evaluated using programs such as \textsc{Matlab}\textsuperscript{\textregistered} with built-in \emph{sparse} matrix exponentiation capabilities. In this way, one solves eq.~\eqref{eq:wtildesol} for a discrete set of $k$-values and then calculates the final result for $c(\mathbf{r},t)$ via a discrete version of eq.~\eqref{eq:cint},

\begin{equation}
c(x,\mathbf{r}_\perp,t) = \delta k \sum_l \tilde{w}(k_l,\mathbf{r}_\perp, t) \exp(i k_l x - D k_l^2 t) 
\label{eq:cintdisc}
\end{equation}

\noindent where the $k_l$ are evenly spaced discrete values of $k$ (with spacing $\delta k$), and the summation runs over a suitably chosen finite set of $k$-values. In the following, we will demonstrate this approach on three important examples.

\section{Shear flow}

First, we consider a linear shear flow above an impenetrable planar surface. The flow direction is the $x$-axis and the $y$-axis is oriented perpendicular to the surface. We consider initial conditions that are independent of the third axis ($z$-axis), so that the solution will also be independent of $z$. We renormalize the coordinates in such a way that the flow profile has the simple form $v(y)=y$. As initial concentration distribution we consider a symmetric Gaussian distribution, 

\begin{align}
    c_0(\mathbf{r},t) = \exp\left(-\frac{\vert \mathbf{r} - \mathbf{r}_0 \vert^2}{2\sigma^2} \right)
    \label{eq:c0}
\end{align}

\noindent with $\mathbf{r}_0=\left\{0,4\right\}$ and $\sigma=1$. The $\{x,y\}$-domain is discretized into square finite elements with edge length 0.1 within the range $-20 < x < 80$ and $0 < y < 20$. Two different boundary conditions are considered at $y=0$: A reflecting boundary, i.e. $c'(x,y=0,t) = \tilde{w}'(k,y=0,t)$, where a prime denotes differentiation after $y$; and a perfectly absorbing boundary, i.e. $c(x,y=0,t) = \tilde{w}(k,y=0,t) = 0$. For all other domain boundaries, we kept the concentration zero for all times (which is, of course, physically acceptable only for sufficiently small integration times). For each considered time value, we first find the discrete vector $\mathbf{\tilde{w}}$ via eq.~\eqref{eq:wtildesol} for 200 discrete values of $k$ given by $k_l = l \Delta k/200-0.5$, $1 \leq l \leq 200$, with $\Delta k = \sqrt{20/(0.1+t)}$. The number and spacing of the discrete $k$-values were chosen in such a way that at zero time the absolute difference between the exact initial concentration distribution $c_0$, eq.~\eqref{eq:c0}, and its discrete representation as given by eq.~\eqref{eq:cintdisc} was everywhere less than $2\cdot10^{-15}$ over the whole considered spatial domain. 

\begin{figure}
\centering\includegraphics[keepaspectratio, width=8cm]{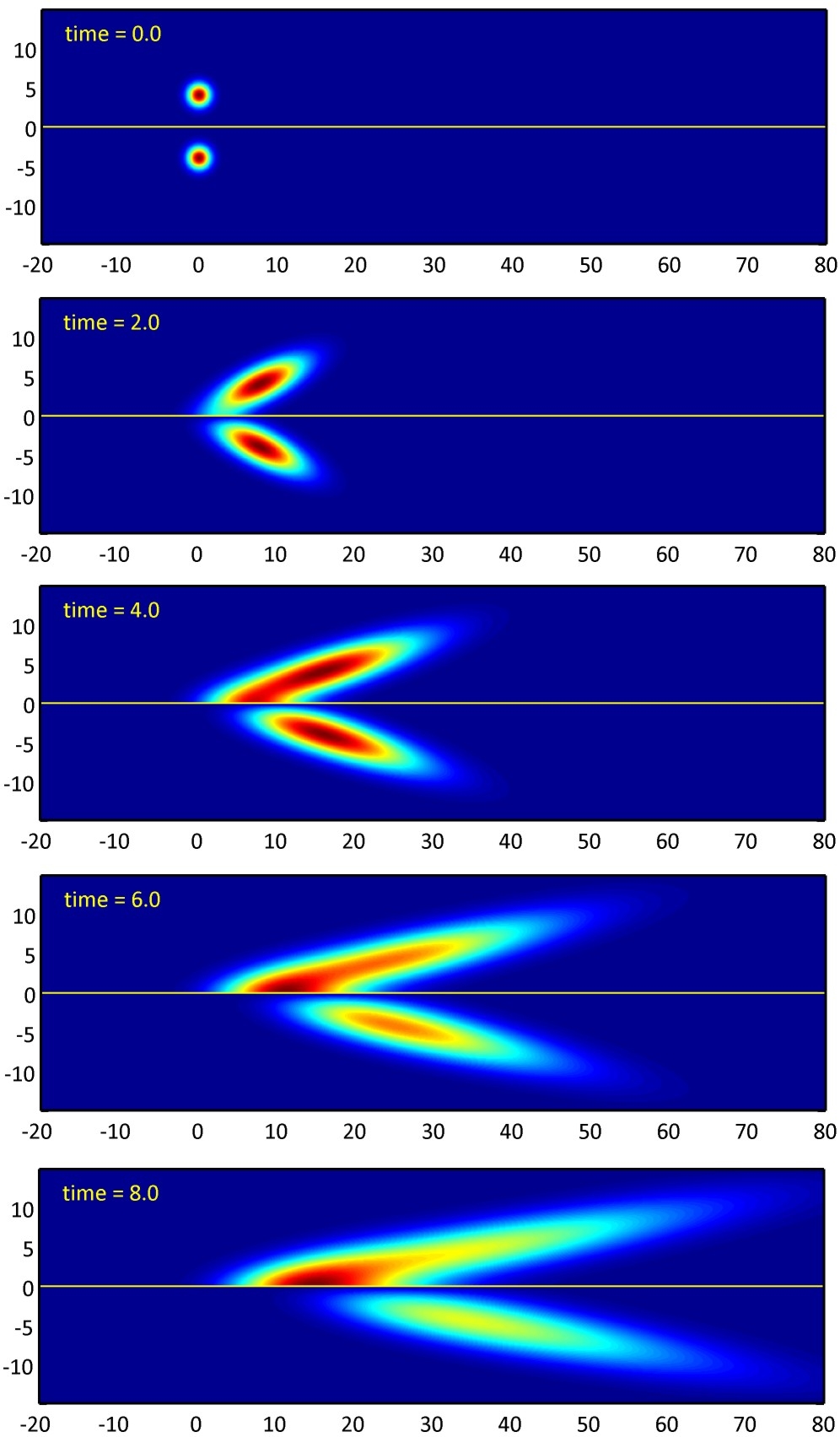}
\caption{Numerical results for the solution of the diffusion-convection for an initially ($t=0$) Gaussian-distributed concentration. Horizontal axis is the $x$-axis, vertical axis is the $y$-axis. Flow direction is from left to right, with flow profile $v=y$. The numerical value of the diffusion coefficient $D$ is one. Upper half planes show the results for a reflecting surface, lower half planes for a perfectly absorbing surface.}
\label{fig:Shear}
\end{figure}

\begin{figure}
\centering\includegraphics[keepaspectratio, width=8cm]{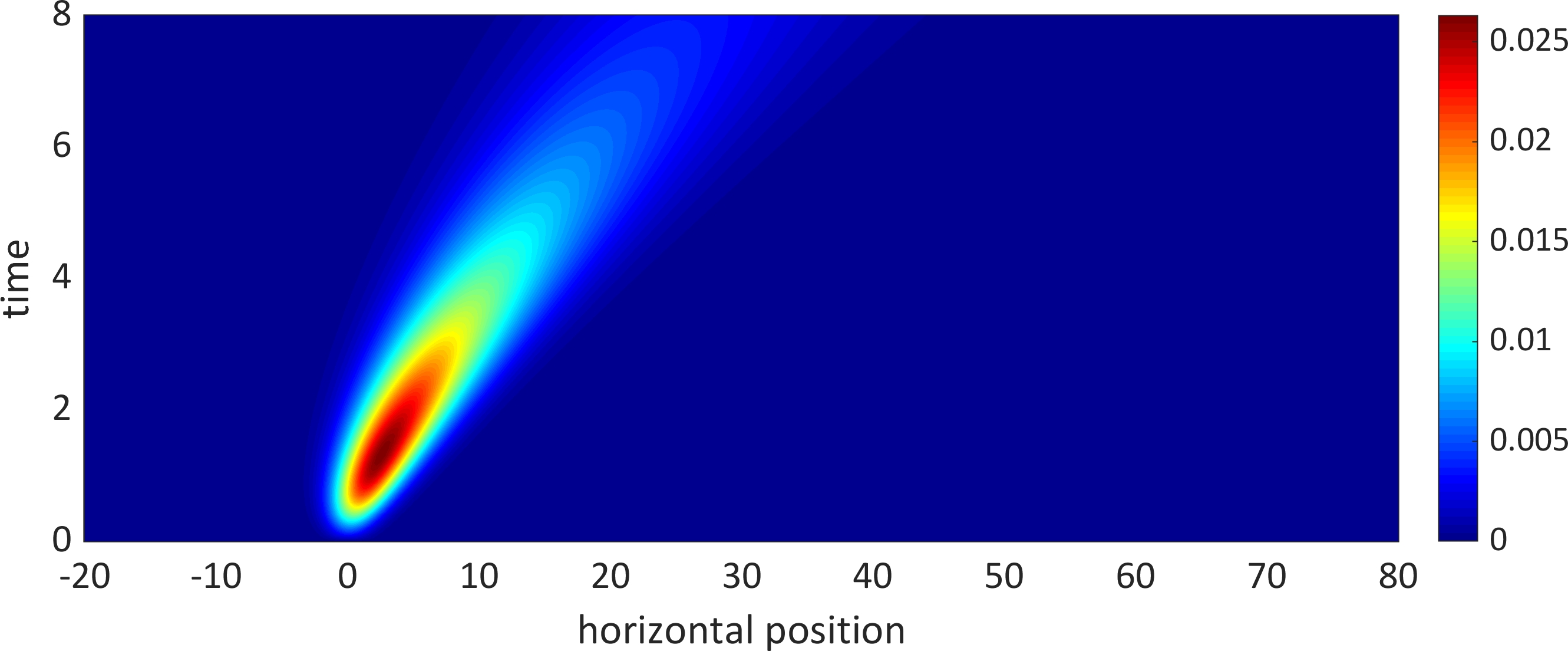}
\caption{Absorption rate $-[\partial c(x,y=0,t)/\partial y]/C_0$ as a function of lateral coordinate $x$ and time $t$ for the solution shown in the bottom half planes of Fig.~\ref{fig:Shear}.}
\label{fig:absrate}
\end{figure}

After having found $\mathbf{\tilde{w}}$, we calculate $c(\mathbf{r},t)$ using eq.~\eqref{eq:cintdisc} with the index $l$ running from $-200$ to 200 and using the fact that due to the special initial conditions we have $\tilde{w}(-k, \mathbf{r}_\perp, t)=\tilde{w}(k, \mathbf{r}_\perp, t)$. The computational results for both reflecting and absorbing boundary conditions for time values $t=2,4,6,8$ are shown in Fig.~\ref{fig:Shear}. One computation for one time value takes less than three seconds on a state-of-the-art desktop computer (Intel\textsuperscript{\textregistered} Xenon\textsuperscript{\textregistered} CPU ES-1650 v4 @ 3.6 GHz). Using the computed result for the absorbing boundary conditions allows us also to compute the relative absorption rate, $-[\partial c(x,y=0,t)/\partial y]/C_0$, as a function of lateral coordinate $x$ and time $t$, where $C_0=\int d\mathbf{r} c_0(\mathbf{r})$ is the initial total amount of matter. This relative absorption rate for the diffusion-convection process from Fig.~\ref{fig:Shear} is shown in Fig.~\ref{fig:absrate}. 

Although we have considered here only purely absorbing or purely reflecting boundary conditions, the presented approach will also work for mixed boundary conditions of the form $\partial c(x,y,t)/\partial y + b c(x,y,t) = d$ with arbitrary constants $a, b$ and $d$. The same condition should then be simply used when solving for the functions $\tilde{w}(k,y,t)$.

\section{Taylor-Aris dispersion}

\begin{figure}
\centering\includegraphics[keepaspectratio, width=8cm]{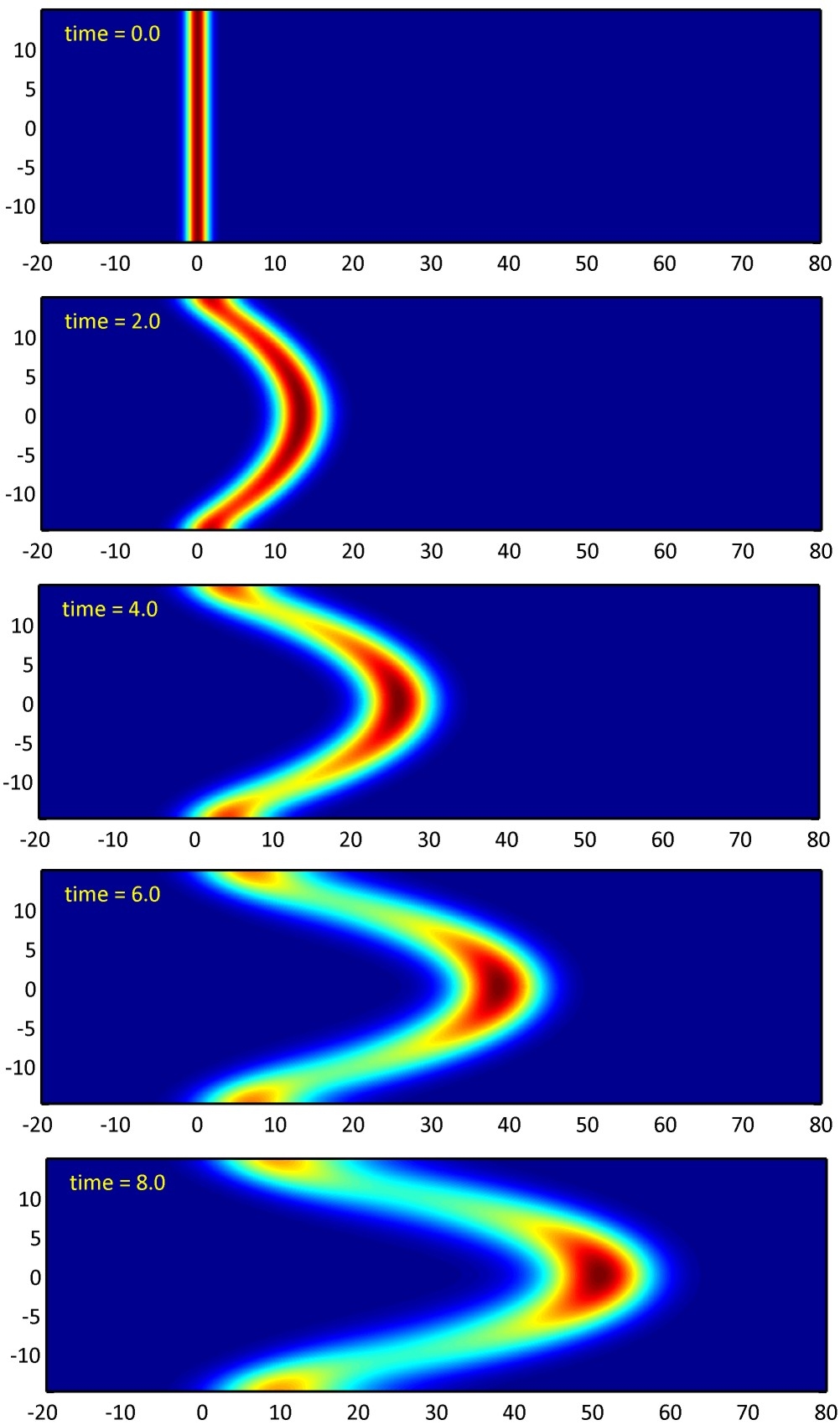}
\caption{\label{fig:TaylorAris} Numerical results for the solution of the diffusion-convection equation for a parabolic flow in a cylindrical capillary of radius $a=15$ with an initially Gaussian-distributed concentration along the horizontal $x$-axis. The vertical axis is the radial coordinate $r$ (negative and positive values are identical radial positions). Flow direction is from left to right, mean flow speed $\bar{v}$ is 3.375, and the numerical value of the diffusion coefficient $D$ is again one.}
\end{figure}

Taylor-Aris dispersion \cite{taylor1953dispersion,taylor1954dispersion,aris1956dispersion} is an important diffusion-convection effect of great importance in microfluidics \cite{stone2001microfluidics,squires2005microfluidics}. It relates to the enhanced diffusive spreading of matter in a flow with an inhomogeneous flow profile. The classical example is the diffusion-convection in a cylindrical capillary with parabolic flow profile $v(y)$, 

\begin{align}
    v(y) = 2 \bar{v} \left(1-\frac{y^2}{a^2}\right),
    \label{eq:parabolic}
\end{align}

\noindent where $y$ is now the radial coordinate (distance from capillary axis), $\bar{v}$ the mean flow velocity, and $a$ the capillary radius. For numerical purposes, it is best to write the cylindrical diffusion-convection equation in the form

\begin{align}
\begin{split}
&\frac{\partial (\sqrt{y} c)}{\partial t} = \\
&\qquad\left[ D\left(\frac{\partial^2}{\partial x^2} + \frac{\partial^2}{\partial y^2} + \frac{1}{4y^2}\right) - v(y) \frac{\partial}{\partial x} \right](\sqrt{y} c)
\end{split}
\end{align}

\noindent so that the second-order differential operator on the r.h.s looks similar to the Laplace operator in Cartesian coordinates, which is most convenient for discretization. The boundary conditions are: $\partial_y c(x,y)=0$ at $y=a$, the capillary's radius; $\partial_y y c(x,y)=0$ at $y=0$; and at the left and right boundary of the integration domain, we keep the concentration zero. Now, we can apply the same approach as described before, with the only difference that in the exponential on the r.h.s. of eq.~\eqref{eq:wtildesol} appears an additional diagonal matrix with elements $Dt/4y^2\equiv -Dt y^{-1/2} \partial_y^2 y^{1/2}$, where we have used the last expression for discretization. For numerical integration, we discretize the $\{x,y\}$-domain into square elements of edge length 0.1 within the range $-20 < x < 80$ and $0 < y < 15$. As initial concentration distribution we consider a transversally homogeneous Gaussian distribution with unit variance, i.e. $c_0(x,y)=\exp(-x^2/2)$. For each time value $t$, we first use again eq.~\eqref{eq:wtildesol} for finding the vectors $\tilde{\mathbf{w}}$ for a discrete set of $k$-values, and then compute the final solution $c(x,y,t)$ via eq.~\eqref{eq:cintdisc}. For the actual example, we use the same set of discrete $k$-values as in the previous shear flow example. And as before, the calculation of the concentration distribution for one time value takes less than 2.5 seconds. The computational results for four time values $t=2,4,6,8$ are shown in Fig.~\ref{fig:TaylorAris}. 

\begin{figure}[hbt!]
\vspace{0.5cm}
\centering\includegraphics[keepaspectratio, width=8cm]{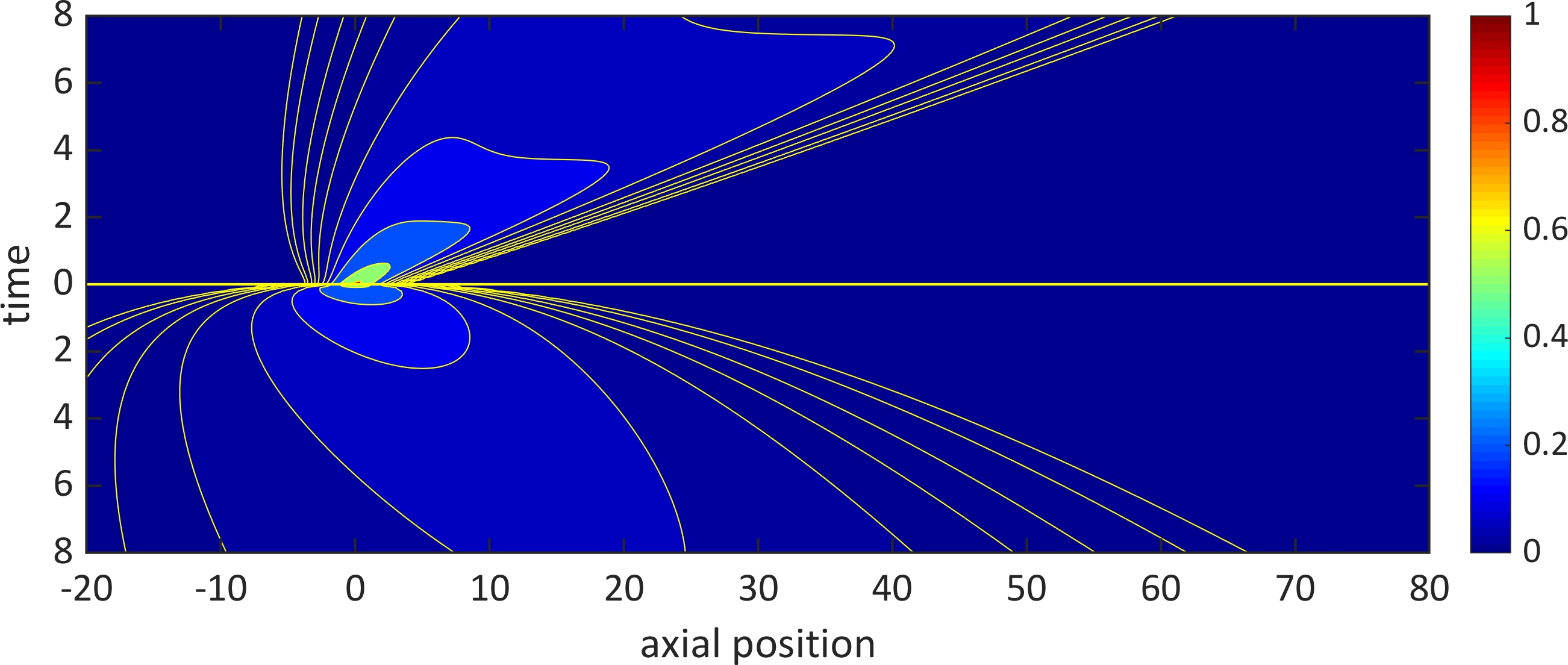}
\caption{Axial dispersion of matter along the capillary's axis. Shown is the concentration integrated over the cross section, $C(x,t)=2 \pi \int dy y c(x,y,t)$. The top half of the image shows the exact result obtained from the numerical solution of the convection-diffusion equation, whereas the bottom half of the image shows the result of the Taylor-Aris approximation. Contour lines are at values 0.001, 0.002, 0.005, 0.01, 0.02, 0.05, 0.1, 0.2 and 0.5.}
\label{fig:DispersionTaylorAris} \end{figure}

Taylor and Aris worked out an approximate solution for the concentration profile along the $x$-direction averaged over the capillary's cross-section. For our initial distribution, this solution reads

\begin{align}
    C(x,t) = \frac{1}{\sqrt{2 D_\textit{eff} t+1}} \exp\left[-\frac{(x-\bar{v}t)^2}{4 D_\textit{eff} t + 2}\right]
    \label{eq:tayloraris}
\end{align}

\noindent where $C(x,t)$ is the concentration integrated over the capillary's cross section, and $D_\textit{eff}$ is an efficient diffusion coefficient given by 

\begin{align}
    D_\textit{eff} = D + \frac{\bar{v}^2 a^2}{48 D} = D \left( 1+\frac{\text{Pe}^2}{192} \right),
    \label{eq:Deff}
\end{align}

\noindent where $\text{Pe}=2 \bar{v} a/D$ is the P\'eclet number for the diffusion in the capillary flow. A comparison between this approximate result and the exact one as calculated from our numerical solution of the convection-diffusion equation is shown in Fig.~\ref{fig:DispersionTaylorAris}. The approximate result (bottom half of figure) strongly differs from the exact numerical result (top figure), which is to be expected for the large value of the P\'eclet number ($\text{Pe} \sim 100$) in our example. 

\begin{figure}
\centering\includegraphics[keepaspectratio, width=8cm]{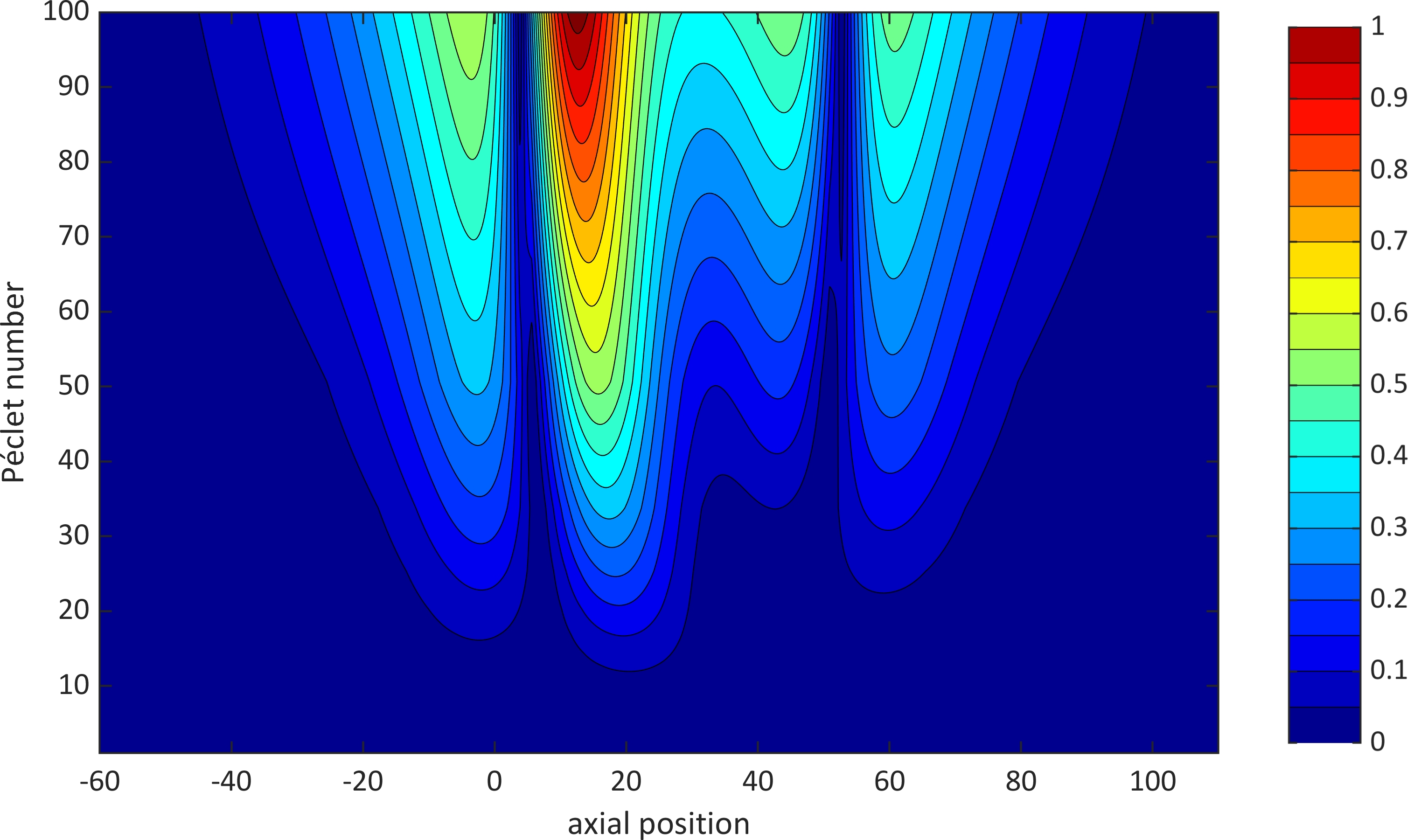}
\caption{\label{fig:TaylorArisPeclet} Difference between numerical simulation result and Taylor-Aris approximation as a function of P\'eclet number. Shown is the relative difference between the numerically computed concentration profile (averaged over the capillary's cross-section) and the Taylor-Aris approximation $C(x,t)$, see eqs.~\ref{eq:tayloraris} and \ref{eq:Deff}. This is done by calculating the absolute difference between both solutions and dividing it by the maximum value of $C(x,t)$. Thus, a value of one signifies that the difference between computed and approximated solution is 100\% the maximum value of the approximation.}  
\end{figure}

To further study the validity of the Taylor-Aris approximation in dependence on the P\'eclet number, we repeated the calculations for P\'eclet number values between 1 and 100 by varying the diffusion coefficient between 1 and 100 in steps of one. To make sure that even for large diffusion coefficient values, the limited integration domain along the $x$-axis does not introduce artifacts, we extended the integration domain no to $-200 \leq x \leq 250$, keeping the grid spacing at 0.1. We calculated the result for the time $t=8$, for the same initial conditions as before. One calculation now took $\sim$20~s. From the full numerical result, we again calculated the concentration profile averaged over the capillary's cross section. This was then compared with with the Taylor-Aris approximation, as shown in Fig.~\ref{fig:TaylorArisPeclet}. In the limit of large diffusion coefficient values (i.e. small P\'eclet numbers), the numerical result becomes indistinguishable from the Taylor-Aris approximation, but starting with P\'eclet number values around 10, the exact result increasingly deviates from the approximation.  

\section{Taylor-Couette laminar flow}

As a third example, we studied the diffusion-convection equation for a laminar Taylor-Couette flow. In particular, we consider the fluid flow between two concentric cylinders of radius $r_1$ and $r_2$, rotating with angular speed $\omega_1$ and $\omega-2$, respectively. If the rotation speeds are not too large, a laminar flow profile will emerge where the flow speed is everywhere directed along the angular coordinate $\phi$ and depends only on the radial coordinate $r$. More precisely, the flow speed as a function of $r$ has the form

\begin{align}
    v(r) = A r + \frac{B}{r}
    \label{eq:taylorcouetteflow}
\end{align}

\noindent with the constants $A$ and $B$ given by

\begin{align}
    A &= \frac{\omega_2 r_2^2 - \omega_1 r_1^2}{r_2^2 - r_1^2}\\
    B &= r_1^2 r_2^2 \frac{\omega_1-\omega_2}{r_2^2 - r_1^2}
    \label{eq:taylorcouettecoeff}
\end{align}

\begin{figure*}
\centering\includegraphics[keepaspectratio, width=\textwidth]{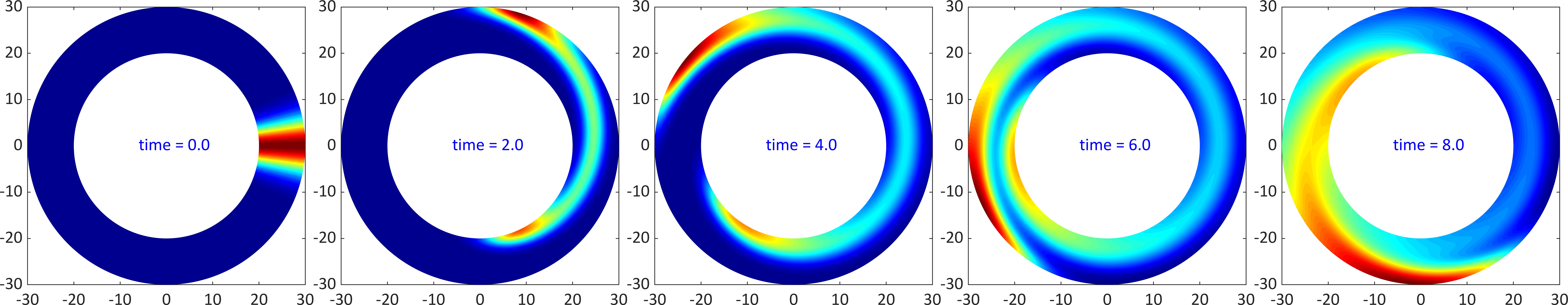}
\caption{\label{fig:TaylorCouette} Numerical results for the solution of the diffusion-convection equation for a laminar Taylor-Couette flow between two cylinders with radius values $r_1=20$ and $r_2=30$. Both cylinders perform a full turn within 8 units of time, the inner cylinder clockwise and the outer counter-clockwise. The numerical value of the diffusion coefficient $D$ is two.}
\end{figure*}

Now, the convection-diffusion equation for the concentration $c(r,\phi,t)$ reads

\begin{align}
\begin{split}
&\frac{\partial (\sqrt{r} c)}{\partial t} = \\
&\left[ D\left(\frac{\partial^2}{\partial r^2} + \frac{1}{r^2}\frac{\partial^2}{\partial \phi^2} + \frac{1}{4r^2}\right) - \frac{v(r)}{r} \frac{\partial}{\partial \phi} \right](\sqrt{r} c)
\end{split}
\end{align}

\noindent where we have used again a similar representation of the Laplace operator as in the last section. The boundary conditions now are $\partial c / \partial r = 0$ at the boundaries $r=r_1$ and $r=r_2$. For the numerical integration, we used the following numerical parameter values: inner cylinder radius $r_1=20$, outer cylinder radius $r_2=30$, angular velocity of inner cylinder $\omega_1=-2\pi/8$, and angular velocity of outer radius $\omega_2=2\pi/8$. Thus, we consider two counter-rotating cylinders which turn exactly once during the full integration time which was again set to $t=8$. We discretized the $\{r,\phi\}$ domain into square elements of edge length 0.02 along $r$ and $\pi/400$ along $\phi$. The initial concentration $c_0$ was chosen to be $\left[\left(1+\cos\phi\right)/2\right]^{100}$, as shown in the leftmost panel in Fig.~\ref{fig:TaylorCouette}. Now, the concentration $c(r,\phi,t)$ is expanded into a Fourier series with respect to the coordinate $\phi$, 

\begin{align}
    c(r,\phi,t) = \sum_{m=-100}^{100} \tilde{w}(r,m,t) \exp(i m \phi)
\end{align}

\noindent For our special choice of the initial concentration $c_0$, this expansion is exact because there are no Fourier components with an absolute value of $m$ larger than 100. For each of the 201 different functions $\tilde{w}(r,m,t)$, the complex-valued diffusion equation that has now to be solved reads

\begin{align}
\begin{split}
&\frac{\partial (\sqrt{r} \tilde{w})}{\partial t} = \\
&\left\{ D\left[\frac{\partial^2}{\partial r^2} -\left(m^2-\frac{1}{4}\right)\frac{1}{r^2}\right]  - \frac{i m v(r)}{r} \right\}(\sqrt{r} \tilde{w})
\end{split}
\end{align}

\noindent which is again done using a similar operator discretization and matrix exponentiation as described in the previous sections. The final numerical results are shown in Fig.~\ref{fig:TaylorCouette}, for integration times of 2, 4, 6 and 8. 

\section{Discussion and Conclusion}

In summary, we have presented a numerical solver for a special but important class of convection-diffusion problems for which the flow velocity does not depend on the coordinate along the flow direction. The importance of our approach is its extreme efficiency: using a standard desktop computer, one can solve complex diffusion-convection problems within seconds, which becomes particularly important when fitting parameters such as flow characteristics of diffusion coefficients against experimental data which requires the frequent repetition of computations for varying parameter sets. 

An important question is the accuracy and stability of our method. There are two aspects to consider. Firstly, we represent the concentration $c(\mathbf{r},t)$ via a Fourier representation, eq.~\ref{eq:cint}, along the coordinate of the flow direction, involving the amplitudes $\tilde{w}(\mathbf{r}_\perp,k,t)$. This is numerically done using a discrete number of Fourier variable values $k$. For an initial concentration distribution such as the Gaussian distributions used for the shear flow and Taylor-Aris examples, such a sum over discrete values of $k$ is only an approximation, and we took great care to use a sufficiently large and suitably spaced number of $k$-values which, for a given initial concentration distribution, approximates it sufficiently well. For the last example, the Taylor-Couette flow, this Fourier representation was even exact, because the given initial concentration could be exactly represented by a finite Fourier sum. Because the partial differential equations for the amplitudes $\tilde{w}(\mathbf{r}_\perp,k,t)$ are linear and all decoupled, no other non-zero Fourier components can occur even for times $t>0$, and the Fourier representation stays exact for all times $t$. Secondly, we have to consider the stability and accuracy of the integration of the differential equations (\ref{eq:wtilde}) for the amplitudes $\tilde{w}(\mathbf{r}_\perp,k,t)$. For this purpose, we use sparse matrix exponentiation as provided by advanced computational software programs such as \textsc{Matlab}\textsuperscript{\textregistered}, see eq.~(\ref{eq:wtildesol}). Because the operator in the exponent is strictly parabolic, this matrix exponentiation is absolutely stable, and its accuracy only depends on its particular numerical implementation within the used software. Of course, it is important to chose a sufficiently fine discretization along the coordinate $\mathbf{r}_\perp$ to assure a sufficiently well approximation of the continuous functions $\tilde{w}(\mathbf{r}_\perp,k,t)$. An important check is whether the numerical result preserves the total concentration over the whole integration domain, and within the limits of our numerical computer precision, this was indeed the case for all the examples presented. It should be also emphasized that a matrix exponentiation approach as used in eq.~(\ref{eq:wtildesol}) does not accumulate numerical errors as in more conventional sequential step-wise integration schemes. For a detailed discussion of the accuracy of numerical matrix exponentiation as implemented in \textsc{Matlab}\textsuperscript{\textregistered}, we refer the reader to the vast technical literature concerning this and similar advanced software. 

An interesting extension of our method is its generalization to three dimensions. This can be done in a straightforward manner as long as the flow velocity profile does not change along the flow direction. We will consider this extension to three dimensions in a future publication which will be devoted to diffusion-convection in straight microfluidic channels with non-trivial cross-sections.    

In the future, it will be interesting to study whether our approach can be generalized to non-Cartesian geometries, for example to a flow along curved pipes. At the moment, our approach is restricted to geometries where we can split the Laplace operator into a transversal part independent on the coordinate of the flow direction, and the remaining part along the flow direction.

\acknowledgments{
Financial support by the Deutsche Forschungsgemeinschaft is gratefully acknowledged (SFB 937, project A5). We thank Marcus M\"uller for many fruitful discussions.


\end{document}